\documentclass[11pt]{article}
\usepackage{epsfig}
\newcommand{\be}{\begin{equation}}
\newcommand{\ee}{\end{equation}}
\newcommand{\ba}{\begin{eqnarray}}
\newcommand{\ea}{\end{eqnarray}}
\newcommand{\no}{\nonumber \\}
\newcommand{\vlk}{ {\rm V}_{\rm low-k}}

\begin{document}
\begin{flushright}
USC(NT)-05-03 : \today
\end{flushright}

\begin{center} 

{\Large\bf Use of $\vlk$ in a 
 chiral-perturbation-theory description of 
 the $pp\rightarrow pp\pi^0$ reaction}

\vskip 5mm 

\large{
Y. Kim$^{(a)}$\footnote{E-mail:ykim@physics.sc.edu}, 
I. Danchev$^{(a)}$\footnote{E-mail:danchev@physics.sc.edu}, 
K. Kubodera$^{(a)}$\footnote{E-mail:kubodera@sc.edu}, 
F. Myhrer$^{(a)}$\footnote{E-mail:myhrer@sc.edu} 
T. Sato$^{(b)}$\footnote{E-mail:tsato@phys.sci.osaka-u.ac.jp}
}

\vskip 5mm 

{\it 
${}^{(a)}$ 
Department of Physics and Astronomy, \\ 
University of South Carolina,
Columbia, 
SC 29208, USA \\ 
${}^{(b)}$
Department of Physics, Osaka University, Toyonaka, Osaka 560, Japan 
}
\end{center} 

\vskip 5mm 

Previously, the near-threshold 
$pp\rightarrow pp\pi^0$ reaction was studied
with the use of transition operators
derived from chiral perturbation theory ($\chi$PT)
and the nuclear wave functions generated
by high-precision phenomenological potentials. 
A conceptual problem in that approach
was that the transition amplitude receives
contributions from very high momentum components 
(above the cutoff scale of $\chi$PT)
in the nuclear wave functions.
In the present work, we avoid this problem
by replacing the ``bare" phenomenological potentials
with $\vlk$, which is an effective potential
derived from a bare potential
by integrating out momentum components 
higher than a specified cutoff scale.
The use of $\vlk$ is found to give an
enhancement of the $pp\rightarrow pp\pi^0$
cross sections over the values obtained with bare potentials.
Although this enhancement brings 
the calculated cross sections
closer to the experimental values,
the incident-energy dependence of the cross 
section is not well reproduced,
a problem that seems to indicate the necessity 
of including higher chiral order terms than considered
in the present work.

\newpage 

\renewcommand{\thefootnote}{\arabic{footnote}}
\setcounter{footnote}{0} 

\section{Introduction}

There have been many theoretical 
investigations~\cite{lr93}-\cite{Hanhart04}
devoted to explaining the 
high-precision data for
the total cross section of
the near-threshold
$pp\rightarrow pp\pi^0$ 
reaction~\cite{meyetal90,Uppsala}. 
The initial surprise was
that the measured cross section
was larger than the values expected
from the earlier calculations
~\cite{kr66,ms91}
by a factor of $\sim 5$. 
Calculations in the phenomenological 
one-boson-exchange model 
indicated that heavy-meson ($\sigma$ and $\omega$) 
exchange contributions could account for 
the unexpectedly large cross section for 
$pp\rightarrow pp\pi^0$~\cite{lr93}. 
The importance of heavy-meson exchanges 
in $\pi^0$ production
is to be contrasted with their much less pronounced
role in the charged-pion production process,
which is dominated by the one-pion-exchange diagrams.
Effective field theory (EFT),
or more specifically, 
chiral perturbation theory 
($\chi$PT) offers a systematic framework
for describing the $NN\rightarrow NN\pi$ processes
at low energies.
The leading-order term in $\chi$PT
(the Weinberg-Tomozawa term)
contributes to charged-pion production
but not to $\pi^0$ production.
$\chi$PT allows us to keep track of
the contributions of higher chiral-order terms 
to the low-energy $NN\rightarrow NN\pi$ 
reactions~\cite{pmmmk96,cfmv96}. 
A point to be kept in mind, however, is
that the $NN\rightarrow NN\pi$ processes
involve rather large momentum transfers,
$p \sim \sqrt{m_N m_\pi}$ 
($m_N$ = nucleon mass, 
$m_\pi$ = pion mass) even at threshold,
and that this feature leads to 
the relatively slow convergence of 
the $\chi$PT expansion~\cite{kmr96}. 
The existence of the additional scale 
$p \sim \sqrt{m_N m_\pi}$ 
in the $NN\rightarrow NN\pi$ reaction led 
Cohen et al.\cite{cfmv96,hkm00} to 
propose a new counting scheme in 
which the expansion parameter is 
$\chi \equiv \sqrt{m_\pi/m_N}$ instead of 
$\chi_W \equiv m_\pi/m_N$ employed in the usual 
Weinberg counting scheme. 
A thorough discussion on this and related topics as
well as an extensive list of references can be found in 
a recent review by Hanhart~\cite{Hanhart04}.  

\vspace{3mm} 

To maintain formal consistency 
in the $\chi$PT calculation of an inelastic nuclear process,
one should derive from the same effective Lagrangian
the relevant transition operators 
and the wave functions for the initial and final 
nuclear states. 
This type of calculation, however, has not yet
been carried out. 
A practical and, in many cases, very useful
method is a hybrid $\chi$PT 
approach~\cite{pmr93}-\cite{KP04},
in which the transition operators are derived
from $\chi$PT but the nuclear wave functions 
are generated with the use of
a modern high-precision phenomenological N-N potential.
Hybrid $\chi$PT was applied 
to the $pp\rightarrow pp\pi^0$ reaction
in Refs.~\cite{pmmmk96}-\cite{apm01}.
These studies indicated: 
(1) There is a substantial cancellation
between the one-body impulse approximation (IA)
term and the two-body contributions,
resulting in a cross section that is
much smaller than the experimental value;
(2) This feature seems reasonably stable against 
the different choices of phenomenological
NN-potentials.\footnote{Ref.~\cite{SLMK97}
reports that
the two different N-N potentials,  
Argonne V18~\cite{wir95} and 
Reid soft-core~\cite{reid},
give almost the same results for the
$pp\rightarrow pp\pi^0$ cross sections.} 

\vspace{3mm} 

A conceptual problem one encounters
in these hybrid $\chi$PT calculations
is that, whereas the transition operators  
are derived using $\chi$PT with the assumption 
that relevant momenta are sufficiently small
compared with the chiral scale 
$\Lambda_\chi$
($p \ll \Lambda_\chi\approx 1$ GeV), 
the wave functions generated by a
phenomenological N-N potential 
can in principle contain momenta 
of any magnitude. 
A numerical calculation 
in Ref.~\cite{SLMK97} indicates 
that the transition amplitude receives non-negligible 
contributions from momentum components 
well above $\Lambda_\chi$,
a feature that jeopardizes 
the applicability of $\chi$PT.

\vspace{3mm}

In a version of hybrid $\chi$PT 
called EFT$^*$ or 
MEEFT~\cite{PKMRa}-\cite{Aetal}, 
the contribution of the 
dangerously high momentum components in  
the wave functions
are suppressed by attaching 
a momentum cutoff factor 
to the transition operators derived from 
$\chi$PT.\footnote{
Another important aspect 
of EFT* is that the low-energy constants 
appearing in the theory are constrained
by the experimental data for the observables
involving neighboring nuclei.
This aspect of EFT*, however,
will not be discussed here.}
EFT* has proved to be extremely useful
in explaining and predicting 
many important observables for electroweak processes
in few-nucleon systems.
Another possible way to suppress 
the contributions of high momentum components
in hybrid $\chi$PT calculations
is to attach a momentum cutoff factor to
the wave functions.\footnote{
Insofar as the use
of a momentum cutoff factor
can be identified with the introduction
of a projection operator onto a model space  
with a limited momentum range,
applying the cutoff factor 
to the transition operators 
is equivalent to applying it 
to the wave functions.}
Meanwhile, a systematic method was developed
by the Stony Brook group and others
~\cite{vlk01,vlk03}
to construct from a phenomenological NN-potential
an {\it effective} NN potential
that resides in a model space which 
only contains momentum components 
below a specified cutoff scale $\Lambda$.
This effective potential, referred to
as $\vlk$, 
is obtained by integrating out 
momentum components higher than $\Lambda$
from a phenomenological NN-potential,
which in this context may be regarded as
an underlying  ``bare" potential
that resides in full momentum space.
$\vlk$ represents
a renormalization-group-improved effective 
interaction of a bare NN interaction.
It has been found
that, for a choice of 
$\Lambda \sim 2\;{\rm fm}^{-1}$,
$\vlk$ reproduces low-energy observables 
such as the phase shifts (for $p<\Lambda$)
and the deuteron binding energy 
with accuracy comparable to that
achieved with the use of bare
high-precision phenomenological 
potentials~\cite{vlk03}.
Furthermore, for any choice of 
bare NN-interactions (belonging to 
the category of modern high-precision 
phenomenological potentials),
it has been found that the corresponding $\vlk$
generates practically the
same half-off-shell T-matrix 
elements for $p<\Lambda$.
This means that the low-momentum
behavior ($p<\Lambda$) of the two-nucleon
wave functions calculated from $\vlk$
is essentially model-independent.

\vspace{3mm}

These developments motivate us
to carry out a hybrid $\chi$PT calculation of
the near-threshold 
$pp\rightarrow pp\pi^0$ reaction
with the use of $\vlk$.
This type of calculation will
substantially reduce the severity of 
the conceptual problem 
of momentum component mismatching 
that existed in the 
previous hybrid $\chi$PT calculations.
It will thus allow us to examine 
more directly whether the transition operators 
derived from $\chi$PT up to a given chiral order
are adequate or not.
Furthermore, 
comparison of the results of a calculation 
based on $\vlk$ with those based on bare
NN-interactions will also give information 
about the influences of the short-distance behavior 
of the NN-interactions on the
$NN\rightarrow NN\pi$ reactions.
In this context, it is informative
to gather more examples of calculations
that use bare NN-interactions.
Therefore, in addition to 
a calculation based on $\vlk$,
we extend here our previous 
bare-potential-based calculations
(carried out for the Argonne V18 and Reid soft-core
potentials) to the Bonn-B potential~\cite{Bonnb}
and the CD Bonn potential~\cite{CDBonn}.

\vspace{3mm}

After describing the primary motivations of our work,
we must mention that our present study 
is basically of exploratory nature and falls short of 
addressing a number of issues that 
warrant detailed studies.
For one thing, we limit ourselves to the use of
the Weinberg counting scheme,
although it is important to examine the consequences
of the counting scheme of Refs.~\cite{cfmv96,hkm00}.
As for the employment of  $\vlk$ in hybrid $\chi$PT,
there is a problem of formal consistency
in that,  whereas the relevant transition operators are
derived in the framework 
of the dimensional regularization~\cite{pmmmk96,SLMK97},
$\vlk$ is based on the momentum cutoff scheme.
Furthermore, the difference between the cutoff scales 
appearing in Ref.~\cite{SLMK97} and 
in $\vlk$ needs to be addressed.
We relegate the study 
of these points to future work
and concentrate on the examination of
the consequences of the use of $\vlk$
in the present limited context; 
on the last point, however, 
we will give some brief remarks later in the text.  

\vspace{3mm}

The organization of this paper is as follows.
Sect. 2 gives a brief recapitulation
of the general framework of hybrid $\chi$PT,
while we explain in Sect.3 some technical aspects 
of numerical calculations we need to address
as we work with $\vlk$ instead of the bare potential.
The numerical results are presented in Sect. 4
and compared with the data.
Finally, Sect. 5 is dedicated to 
discussion and summary.

\section{Calculational Framework}

The formalism to be used here is
basically the same as in Refs.~\cite{pmmmk96, SLMK97}
except that, for a calculation with $\vlk$,
some modifications (essentially of technical nature)
are needed. Therefore, as far as the general framework 
of our approach is concerned, we only give a brief
recapitulation, referring to 
Refs.~\cite{pmmmk96, SLMK97} for details.

\subsection{Transition operators}

As in Refs.~\cite{pmmmk96, SLMK97},
we derive the transition operators
for the $pp\rightarrow pp\pi^0$ reaction
using the heavy-fermion formalism
(HFF)~\cite{jm91} of $\chi$PT
based on the Weinberg counting rules.
The relevant lagrangian is written as
\begin{equation}
{\cal L}_{\rm{ch}} = {\cal L}^{(0)} + {\cal L}^{(1)}
+ {\cal L}^{(2)}+ \cdots\;\; .
\label{eq:Lag0}
\end{equation}
Here ${\cal L}^{(\bar{\nu})}$ 
($\bar{\nu}=0,1,2 \ldots$)
contains terms of chiral order $\bar{\nu}$
with $\bar{\nu}\equiv d + (n/2) - 2$,
where $n$ is the number of
fermion lines involved in the vertex 
and $d$ is the number of derivatives or
powers of $m_\pi$. 
For our present study we only
need the terms of $\bar\nu=0$ and $\bar\nu=1$, 
which are given as follows:   
\begin{eqnarray}
{\cal L}^{(0)} &=&
 \frac{f^2_\pi}{4} \mbox{Tr}
[ \partial_\mu U^\dagger \partial^\mu U
 + m_\pi^2 (U^\dagger +  U - 2) ]
 + \bar{N} ( i v \cdot D + g_A^{} S \cdot u ) N
\label{eq:L0}\\
{\cal L}^{(1)} &=&
-\frac{i g_A^{}}{2m_N} \bar{N}
\{ S \!\cdot\! D, v \!\cdot\! u \} N
 + 2c_1 m_\pi^2 \bar{N} N \mbox{Tr}
( U + U^\dagger - 2 )  \nonumber \\
 &&+ (c_2 \!-\! \frac{g_A^2}{8m_N}) \bar{N}
(v \!\cdot\! u)^2 N
 + c_3 \bar{N} u \!\cdot\! u  N \; . 
\label{eq:L1}
\end{eqnarray}
Here $U(x)$ is an SU(2) matrix
that is non-linearly related to the pion field
and that has standard chiral
transformation properties;
we use 
$U(x) = \sqrt{1-[{\vec \pi}(x)/f_\pi]^2}
+i{\vec\tau} \!\cdot\! {\vec \pi}(x)/f_\pi$.
$N(x)$ denotes the large component of 
the heavy-fermion field;
the four-velocity parameter $v_\mu$
is chosen to be $v_\mu=(1,0,0,0)$.
$D_\mu N$ is the covariant derivative, 
$S_\mu$ is the covariant spin operator, and 
$u_\mu\equiv i [\xi^\dagger \partial_\mu \xi
- \xi \partial_\mu \xi^\dagger]$, where
$\xi = \sqrt{U(x)}$ \cite{bkm95}.
The pion decay constant
is taken to be $f_\pi=93$ MeV, and
$g_A=1.25$.
The values of the low-energy constants (LECs), 
$c_1,c_2$ and $c_3$, 
are given in, e.g.,  
Refs.~\cite{pmmmk96,bkm95,bkm97}:
\begin{equation}
c_1=-0.87\pm 0.11\;{\rm GeV}^{-1},\;\;
c_2=3.34\pm 0.27\;{\rm GeV}^{-1},\;\;
c_3=-5.25\pm 0.22\;{\rm GeV}^{-1}.
\label{eq:lecoef}
\end{equation}

\vspace{3mm} 

The chiral order index $\nu$ of a Feynman diagram
is defined by 
$\nu = 4 - E_N - 2C + 2L + \sum_i \bar{\nu}_i$,
where $E_N$ is the number of nucleons
in the Feynman diagram,
$L$ the number of loops,
$C$ the number of disconnected parts
in the diagram, and the sum runs over all the vertices
in the Feynman graph~\cite{wei90}. 
We are using here the Weinberg counting scheme~\cite{wei90},
although, as mentioned in the introduction, 
there exists a different counting scheme
tailored to keep track of high-momentum flows involved in 
$NN\rightarrow NN\pi$ reactions~\cite{cfmv96,Hanhart04}. 
Furthermore, we limit ourselves to the tree-level diagrams 
and examine the consequences of 
employing $\vlk$ in evaluating the transition
matrix elements corresponding to these tree diagrams.
%
%
\begin{figure}
\centerline{\epsfig{file=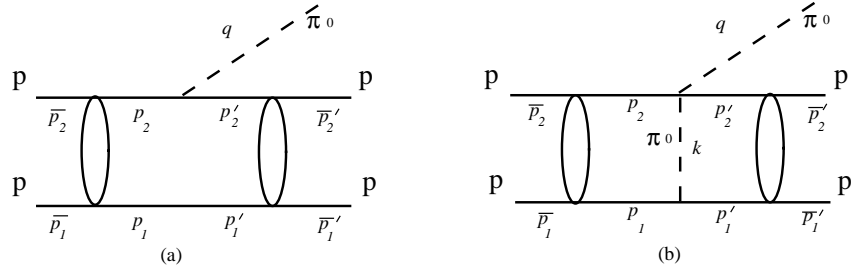,width=11.2cm}} 
\caption{\small
Impulse term (a) and rescattering term 
(b) for the 
$pp\rightarrow pp\pi^0$ reaction. 
In the text, the space components of the initial 
four-momenta $\bar{p}_1$ and $\bar{p}_2$ 
in the center-of-mass system are denoted by
$\vec{p}_i$ and $-\vec{p}_i$, respectively; 
similarly, the space components of the final
four-momenta $\bar{p}_1'$ and $\bar{p}_2'$ 
in the center-of-mass system are denoted by 
$\vec{p}_f-\vec{q}/2$ 
and $-\vec{p}_f-\vec{q}/2$, respectively.
}
\label{impres}
\end{figure}

\vspace{3mm}

The kinematic variables for the
$pp\rightarrow pp\pi^0$ reaction we use
in this work
and the relevant Feynman diagrams 
are shown in Fig. 1.
As discussed in Refs. \cite{pmmmk96, SLMK97},
the lagrangian in Eq.~(\ref{eq:Lag0}) 
leads to the transition operator 
\be
{\cal T} \,=\, {\cal T}^{(-1)}+{\cal T}^{(+1)}
\equiv {\cal T}^{({\rm Imp})} +
       {\cal T}^{({\rm Resc})}\label{eq:Tdecomposed}\,,
\ee
where ${\cal T}^{(\nu)}$ represents the contribution
of chiral order $\nu$.
${\cal T}^{(-1)}$ comes from the one-body 
impulse approximation (IA) diagram  
[Fig. 1(a)] and is given by 
\begin{eqnarray}
{\cal T}^{(-1)}\equiv {\cal T}^{({\rm Imp})}
\equiv \frac{i}{(2\pi)^{3/2}}
\frac{1}{\sqrt{2\omega_q}}\frac{g_A}{2f_\pi}
\sum_{j=1,2} [-\vec{\sigma}_j\cdot\vec{q}
+\frac{\omega_q}{2m_N}\vec{\sigma}_j
\cdot (\vec{p}_j + \vec{p}_j^{\,\,\prime} )]\tau_j^0 \; .
\label{eq:Tminus1} 
\end{eqnarray}
${\cal T}^{(+1)}$, which arises from 
the two-body rescattering diagram [Fig.1(b)],
is given by 
\begin{eqnarray}
{\cal T}^{(+1)}\equiv {\cal T}^{({\rm Resc})}
\equiv \frac{-i}{(2\pi)^{9/2}}
\frac{1}{\sqrt{2\omega_q}}\frac{g_A}{f_\pi}
\sum_{j=1,2} \kappa(k_j,q)\frac{\vec{\sigma}_j
\cdot \vec{k}_j \; \tau_j^0}{k_j^2 - m_\pi^2} \; ,
\label{eq:Tplus1}
\end{eqnarray}
where $\vec{p}_j$ and $\vec{p}^{\,\,\prime}_j$ ($j=1,2$)
denote the initial and final
momenta of the $j$-th proton.
The four-momentum of the exchanged pion is defined by the
nucleon four-momenta at the $\pi NN$ vertex as 
$k_j \equiv p_j-p^{\,\,\prime}_j$, 
where $p_j =(E_{p_j}, \vec{p}_j),
p^{\,\,\prime}_j=
(E_{p^{\,\,\prime}_j}, \vec{p}^{\,\,\prime}_j)$ with
$E_p\equiv(\vec{p}^{\,\,2} + m_N^2)^{1/2}$.
The four-momenta of the final pion
is $q=(\omega_q, \vec{q}\,$),
with $\omega_q=(\vec{q}^{\; 2}+m_\pi^2)^{1/2}$.
The rescattering vertex function
$\kappa(k,q)$ of Eq.(\ref{eq:Tplus1}) 
is calculated from Eq.(\ref{eq:L1}):
\begin{equation}
\kappa(k,q)\equiv \frac{m_\pi^2}{f_\pi^2}\,
[\,2c_1 - (c_2 - \frac{g_A^2}{8m_N})
\frac{\omega_q k_0}{m_\pi^2}
 - c_3 \frac{q\cdot k}{m_\pi^2}\,]\;,
\label{eq:kappakq}
\end{equation}
where $k=(k_0,\vec{k})$ represents the four-momenta
of the exchanged pion. 

\subsection{Transition amplitude and 
nuclear wave functions}

We write the transition amplitude
for the $pp \rightarrow pp\pi^0$ reaction as
\begin{equation}
T\,=\,\langle \Phi_f | {\cal T} | \Phi_i \rangle,
\label{eq:Tmatrix}
\end{equation}
where $|\Phi_i\rangle$ ($|\Phi_f\rangle$)
is the initial (final) two-nucleon state
distorted by the initial-state (final-state) interaction. 
As briefly discussed in the introduction, 
in a formally consistent nuclear
$\chi$PT calculation, 
the transition operator ${\cal T}$
and the N-N interactions that
generate $|\Phi_i\rangle$ and
$|\Phi_f\rangle$ are to be calculated
to the same chiral order $\nu$
from the common $\chi$PT lagrangian.  
In hybrid $\chi$PT,
we instead use a phenomenological N-N potential 
to generate $|\Phi_i\rangle$ and
$|\Phi_f\rangle$.
In the present treatment 
this phenomenological N-N potential
can be either $\vlk$ or a bare NN-interaction
(see the introduction).

\vspace{3mm}

As described in Ref.~\cite{SLMK97},
we can apply the standard partial-wave decomposition
to Eq.~(\ref{eq:Tmatrix}) and rewrite it into
\ba
\langle\chi^{(-)}_{\vec{p}_f,m_{s_1'},m_{s_2'}}
\vec{q}\,|{\cal{T}}
|\chi^{(+)}_{\vec{p}_i,m_{s_1},m_{s_2}}\rangle
&=& \sum_{S_fL_fJ_fM_f}\sum_{S_iL_iJ_iM_i}
{\cal{Y}}_{S_fL_f}^{J_fM_f^+}
(\hat{p_f},m_{s_1'},m_{s_2'})
{\cal{Y}}_{S_iL_i}^{J_iM_i}
(\hat{p_i},m_{s_1},m_{s_2})\nonumber\\
 &\times&\sum_{l_{\pi}m_{\pi}}
Y^*_{l_{\pi}m_{\pi}}(\hat{q})\,
\langle p_f[L_fS_f]J_fM_f|
{\cal{T}}_{l_{\pi}m_{\pi}}(q)|
p_i[L_iS_i]J_iM_i\rangle\,,
\ea
Here ${\cal{Y}}_{SL}^{JM}$ 
is the spin-angular function
of the antisymmetrized two-proton state 
\ba
{\cal{Y}}_{SL}^{JM}\equiv
\frac{1+(-1)^{L+S}}{\sqrt{2}}
\sum_{M_SM_L}i^L\exp[i\delta_{(LS)J}]\,
Y^*_{LM_L}(\hat{p})
\langle \frac{1}{2}\frac{1}{2} m_{s_1}m_{s_2}
|SM_S\rangle
\langle LSM_LM_S|JM\rangle\,,
\ea
where $\delta_{(LS)J}$ is the NN scattering phase shift
in the eigen-channel defined by the orbital
angular momentum $L$, total spin $S$,
and the total angular momentum $J$.
$l_\pi$ denotes the angular momentum of 
the outgoing pion.
It is convenient to introduce the reduced matrix element
using the standard convention:
\ba
\lefteqn{\langle p_f[L_fS_f]J_fM_f|
{\cal{T}}_{l_{\pi}m_{\pi}}(q)|
p_i[L_iS_i]J_iM_i\rangle }\\
&&\equiv (-1)^{J_f-M_f}
\left(
\begin{array}{ccc}
 J_f & l_\pi & J_i\\
-M_f & m_\pi & M_i
\end{array}
\right)
\langle p_f[L_fS_f]J_f\parallel
{\cal{T}}_{l_{\pi}m_{\pi}}(q)\parallel
p_i[L_iS_i]J_i\rangle\,.
\ea
Corresponding to the decomposition
${\cal T}\,=\, 
{\cal T}^{{\rm Imp}}+{\cal T}^{{\rm Resc}}$
in Eq.(\ref{eq:Tdecomposed}),
the reduced matrix element has two terms
\begin{eqnarray}
\lefteqn{<\!p_f[L_fS_f]J_f\,||\,{\cal T}_{l_\pi}(q)
\,||\,p_i [L_iS_i]J_i\!>}\nonumber\\
& &\;\;\;=<\!p_f[L_fS_f]J_f\,||\,
{\cal T}^{\rm Imp}_{l_\pi}(q)
\,||\,p_i [L_iS_i]J_i\!> +
<\!p_f[L_fS_f]J_f\,||\,{\cal T}^{\rm Resc}_{l_\pi}(q)
\,||\,p_i [L_iS_i]J_i\!>\,. 
\label{eq:Treldecomp}
\end{eqnarray}
Near threshold we can assume that 
the $pp\rightarrow pp\pi^0$ reaction
is dominated by s-wave pion production,
with the final $pp$ states in 
the $^1\!S_0$ partial wave;
this implies that we need only
consider the $^3\!P_0$ partial wave
for the initial $pp$ state. 
With these  constraints, 
the reduced matrix elements for the impulse
and rescattering terms are given by
\begin{eqnarray}
\frac{1}{\sqrt{4\pi}}<\!p_f[^1\!S_0]\,||\,
{\cal T}^{\rm Imp}_{l_\pi=0}(q)
\,||\,p_i [^3\!P_0]\!>\; &=&\;
   \frac{-i}{\sqrt{(2\pi)^3 2\omega_q}}
\frac{g_A}{f_{\pi}}
\int \int\frac{d\vec{p}^{\,\,\prime}d\vec{p}}{4\pi}
  R_{^1\!S_0,\; p_f}(p') \nonumber \\
 &\times&  \hat{p} \cdot (- \vec{q} + 
\frac{\omega_q}{m_N}\vec{p}^{\,\,\prime})
 \delta(\vec{p}^{\,\,\prime} - \vec{p} + \vec{q}/2)
  R_{^3\!P_0,\; p_i}(p)~, 
\label{eq:TImppspace} \\
\frac{1}{\sqrt{4\pi}}<p_f[^1\!S_0]\,||\,
{\cal T}_{l_{\pi}=0}^{\rm Resc}(q)\,||\,p_i [^3\!P_0]>
 &=&  \frac{i}{\sqrt{(2\pi)^3 2\omega_q}}
\frac{2 g_A}{f_{\pi}}
\int \int\frac{d\vec{p}^{\,\,\prime}d\vec{p}}{4\pi}
  R_{^1\!S_0,\; p_f}(p') \nonumber \\
&\times& \frac{\kappa(k,q)}{(2\pi)^3}
\frac{\hat{p} \cdot \vec{k}}{ k^2 - m_{\pi}^2}
  R_{^3\!P_0,\; p_i}(p) \; . 
\label{eq:TRescpspace}
\end{eqnarray} 
The radial functions,
$R_{^3\!P_0,\; p_i}(p)$ 
and 
$R_{^1\!S_0,\; p_f}(p')$, 
in Eqs.(\ref{eq:TImppspace}) and (\ref{eq:TRescpspace})  
stand for the
NN relative motion in the initial and final state,
respectively.
To obtain these radial functions,
we first derive the $K$-matrix 
by solving the Lippman-Schwinger equation in momentum
space for a given NN potential; see Ref.~\cite{ht70}. 
The calculated half-off-shell $K$-matrix and 
N-N phase shift $\delta_{(LS)J}$ give 
the corresponding momentum space radial wave function as: 
\begin{eqnarray}
R_{(LS)J,p_0}(p) = i^{-L} \cos(\delta_{(LS)J})
\left[\frac{\delta(p-p_0)}{p_0^2} +
{\cal P} \; \frac{K_{(LS)J}(p,p_0,W)}
{p_0^2/m_N - p^2/m_N}\right] \; . 
\label{eq:Kmatrix}
\end{eqnarray}
Here ${\cal P}$ means
taking the principal-value part of the two-nucleon propagator,
and $p_0$ is the on-shell momentum defined by $W=2E_{p_0}$.
We note that the on-shell $K$-matrix
is related to the phase shift
as 
$\tan (\delta_{(LS)J})
=-\pi p_0 m_N K_{(LS)J}(p_0,p_0)/2$.  

\vspace{3mm} 

The choice of the four-momentum $k$ 
of the exchanged pion has been a subject of
investigations in the literature, 
see e.g., Refs~\cite{pmmmk96,hmmsv01}.
In a simple prescription, 
which came to be known as 
fixed kinematics approximation (FKA)~\cite{pmmmk96},
we identify the four-momentums of the intermediate nucleon lines
with the corresponding asymptotic values
(ignoring thereby energy-momentum transfers
due to the initial and final state interactions)
and ``freeze" all the kinematic variables 
at their threshold values.
Thus FKA consists in using,
in Eqs.(\ref{eq:kappakq}) and (\ref{eq:TRescpspace}),
$k=(k_0,\vec{k})$ = $(m_\pi/2,\vec{k})$, 
$\vec{k}=\vec{p}-\vec{p}^{\,\, \prime} $, and 
$q=(\omega_q,\vec{q})=(m_\pi,\vec{0})$.
Meanwhile, in Ref.~\cite{SLMK97}, 
$k$ was chosen in such a manner that 
energy-momentum conservation at each vertex
in Fig.1(b) should be satisfied, {\it i.e.,}
\begin{equation}
k = (k_0,\vec{k}) = (E_{\vec{p}}-E_{\vec{p}^{\,\,\prime} -
\vec{q}/2}, \vec{p} - \vec{p}^{\,\,\prime} + \vec{q}/2)\,,
\label{eq:kchoicea}
\end{equation}
where 
$\vec p$ ($\vec p^{\; \prime}$) 
is the relative momentum
of the two protons before (after) pion emission, 
defined in 
Fig.~\ref{impres} as:   
\begin{equation}
\vec{p}_1 = - \vec{p}_2 =\vec{p},\;\;\;\;\;\;
\vec{p}_1^{\,\,\prime} =
\vec{p}^{\,\,\prime} -\frac{\vec{q}}{2},\;\;\;\;\;\;
\vec{p}_2^{\,\,\prime} = -\vec{p}^{\,\,\prime}
-\frac{\vec{q}}{2}\,.\label{eq:kchoiceb}
\end{equation}
A schematic study in Ref.~\cite{hmmsv01}
indicates that, when the final-state NN 
interaction is included in the rescattering diagram, 
FKA is an appropriate choice, but that,
when initial-state interaction is included,
the situation is more complex. 
Thus the choice of $k$ is still an open issue.
In the present work therefore
we give numerical results for the choice of $k$
given in Eqs.(\ref{eq:kchoicea}) and (\ref{eq:kchoiceb}), 
as well as for FKA.
The bulk of our calculation 
will be done with the use of $k$ 
given in Eqs.(\ref{eq:kchoicea}) and (\ref{eq:kchoiceb});
the results corresponding to FKA will be presented 
with due remarks attached to them. 

\subsection{Cross sections}

The total cross section at energy $W (=2E_{\vec{p}_i})$ 
in the center-of-mass frame for the 
reaction  $pp\rightarrow pp\pi^0$
is given by~\cite{SLMK97}
\begin{eqnarray}
\sigma_{pp\rightarrow pp\pi^0}(W) & = &
 \frac{(2\pi)^4}{16}\frac{E_{p_i}}{p_i}
\int_{0}^{q_m}\!dq\,q^2 p_f
\sqrt{4E_{p_f}^2 + \vec{q}^{\,2}} \nonumber \\
 & \times &
 \sum_{L_i S_i J_i L_f S_f J_f l_{\pi}}
 | \frac{1}{\sqrt{4\pi}}e^{i\delta_{(L_fS_f)J_f}
+ i\delta_{(L_i,S_i)J_i}} <\!p_f[L_f S_f]J_f\,||\,
{\cal T}_{l_{\pi}}(q)||\,p_i[L_i S_i] J_i\!>|^2\nonumber\\
& & \label{eq:sigmaWb}
\end{eqnarray}
where $E_{p_f}\equiv
\{(W-\omega_q)^2 - \vec{q}^2\}^{1/2}/2$,
$p_f\equiv \sqrt{E_{p_f}^2-m_N^2}$\,,
and the maximum momentum, $q_m$, of the pion
is given by $q_m$ = $\sqrt{\{(W-2m_N)^2-m_\pi^2\}
\{(W+2m_N)^2-m_\pi^2\}/4W^2}$. 
Here $p_i$ ($=|\vec p_i|$) is the asymptotic relative momentum 
of the initial 
$pp$ states and $E_{\vec p_i}= \sqrt{\vec p_i^{\; 2}+m_N^2}$. 
Since we have already specialized ourselves 
in the threshold pion production,
we need not deal with the general expression in
Eq.(\ref{eq:sigmaWb});
we can limit $[L_fS_f]J_f$ to $^1\!S_0$
and $[L_i S_i] J_i$ to $^3\!P_0$.

\subsection{NN interactions}

As discussed in the introduction,
the main purpose of the present work 
is to carry out a hybrid $\chi$PT calculation
of the $pp\rightarrow pp\pi^0$ reaction
with the use of $\vlk$,
which resides in reduced Hilbert space
characterized by the constraint $p<\Lambda$.
Specifically, we use here the $\vlk$
derived from the CD-Bonn potential
~\cite{CDBonn}.\footnote{We are grateful
to T.T.S. Kuo for allowing us to
use a computer code to generate $\vlk$
developed by his group.}
We are also interested in comparing the 
the results of this calculation
with those of hybrid $\chi$PT calculations
based on standard high-precision
phenomenological potentials,
which we refer to as ``bare" interactions.
Regarding a bare NN potential case, 
in order to augment the examples
given in Ref.~\cite{SLMK97},
we shall carry out additional calculations
with the use of the Bonn-B potential
and the CD Bonn potential.

\section{Numerical Calculation}

The numerical evaluation of 
the scattering amplitude and
the cross section follows 
closely the method employed in Ref.~\cite{SLMK97}.
The calculation of the rescattering amplitude
[Eq.~(\ref{eq:TRescpspace})] can be readily 
done in momentum ($p$)-space,
whereas it is technically easier to carry out
a numerical evaluation of 
the impulse amplitude [Eq.(\ref{eq:TImppspace})]
in coordinate ($r$)-space~\cite{SLMK97}.
Since the calculational method for the case of
a bare NN-potential was explained
in detail in Ref.~\cite{SLMK97},
we only describe here
modifications that need to be made
when we use $\vlk$ instead of a bare potential.

\subsection{Principal-value integral}

The principal value integral appearing
in Eq.~(\ref{eq:Kmatrix})
is usually rendered amenable 
to numerical calculation
in the following manner.
If we need to numerically evaluate the integral 
\be
{\rm I}\equiv
{\cal P}\int_0^\infty dk\frac{f(k)}{q^2-k^2}~, 
\ee
we may convert this expression 
into an ordinary integral by subtracting zero: 
\ba
{\rm I}&=&{\cal P}\int_0^\infty dk\frac{f(k)}{q^2-k^2}-
{\cal P}\int_0^\infty dk\frac{f(q)}{q^2-k^2}\no
&=&\int_0^\infty dk\frac{1}{q^2-k^2}[f(k)-f(q)]\, .
\ea
In a calculation that involves $\vlk$,
the upper limit of k-integration is a finite value
($\Lambda$), so that we encounter an integral like
\be
{\rm I}_\Lambda \equiv
{\cal P}\int_0^\Lambda dk\frac{f(k)}{q^2-k^2}~, 
\ee
where, for the sake of definiteness,
we may assume $q < \Lambda$.
Since ${\cal P}\int_0^\Lambda dk
\frac{f(q)}{q^2-k^2}\ne 0$,
the procedure used for ${\rm I}$
needs to be modified as
\ba
{\rm I_\Lambda}&=&
{\cal P}\int_0^\Lambda dk\frac{f(k)}{q^2-k^2}-
{\cal P}\int_0^\Lambda dk\frac{f(q)}{q^2-k^2} 
+{\cal P}\int_0^\Lambda dk\frac{f(q)}{q^2-k^2} \no
&=&\int_0^\Lambda dk\frac{1}{q^2-k^2}[f(k)-f(q)]
+\frac{f(q)}{2q}
\log \frac{\Lambda + q}{\Lambda- q}\, . 
\ea

\subsection{Calculation of 
the impulse-term amplitude in r-space. }

As mentioned, the numerical evaluation 
of the impulse-term amplitude can be
done more conveniently in r-space than
in p-space.
In a case involving a bare potential,
switching from the p-representation 
to the r-representation can be readily performed
using the standard Bessel transformation,
\begin{eqnarray}
R_{(LS)J,p_0}(r) = \sqrt{\frac{2}{\pi}}\,i^L
\!\int_0^\infty \! p^2dp\,j_L(pr)R_{(LS)J,p_0}(p)\,,
\label{eq:Bessel}
\end{eqnarray}
and the well-known identity, 
$\int_0^\infty j_L(pr)j_L(pr^\prime) p^2dp
=\frac{\pi}{2r^2}\delta (r-r^\prime)$.
The result is given by~\cite{SLMK97}
\begin{eqnarray}
\frac{1}{\sqrt{4\pi}}<\!p_f[^1S_0]\,||\,
{\cal T}_{l_{\pi}=0}^{\rm Imp}(q)\,||\,p_i[^3P_0]\!>
  =  \frac{1}{\sqrt{(2\pi)^32\omega_q}}
\frac{g_A}{f_{\pi}}
   \int\!dr\, r^2  R_{^1\!S_0,p_f}(r) &&\nonumber \\
\times \Big[ ( 1 + \frac{\omega_q}{2m_N})
q j_1(qr/2) 
- \frac{\omega_q}{m_N}j_0(qr/2)
(\frac{d}{dr} + \frac{2}{r})\,\Big]\,R_{^3\!P_0,p_i}(r)~.&&
\label{eq:TImprspace}
\end{eqnarray} 

\vspace{3mm} 

The usefulness of this method, 
however, diminishes in the case of $\vlk$, 
where the momentum integral does not run 
to $\infty$ but stops at $\Lambda$,
and hence we cannot use the 
above-quoted orthogonality 
of the spherical Bessel functions:
$\int_0^\Lambda j_L(pr)j_L(pr^\prime) p^2dp
\neq\frac{\pi}{2r^2}\delta (r-r^\prime)$.
We therefore use the following procedure. 
In evaluating the impulse amplitude
[Eq.(\ref{eq:TImppspace})] for  $\vlk$, 
we first integrate out the $\delta$-function,
and then divide the range of
$p$-integration in two intervals as follows. 
\ba
&&\frac{1}{\sqrt{4\pi}}<\!p_f[^1S_0]\,||\,
{\cal T}^{\rm Imp}_{l_\pi=0}(q)
\,||\,p_i [^3P_0]\!>\; =\;
   \frac{-i}{\sqrt{(2\pi)^3 2\omega_q}}
\frac{g_A}{2f_{\pi}}\no
&&\qquad\qquad\qquad\qquad\times\{
 \int_0^{p_c} dpp^2\int_{-1}^{+1} dx
  R_{^1\!S_0,\; p_f}(p) \; 
\Big[\frac{\vec l}{l} \cdot (- \vec{q} + 
\frac{\omega_q}{m_N}\vec{p}\;)\Big] 
  R_{^3\!P_0,\; p_i}(l)\no
&&\qquad\qquad\qquad\qquad\qquad +
\int_{p_c}^\Lambda dpp^2\int_{-1}^{+1} dx
  R_{^1\!S_0,\; p_f}(p) \; 
\Big[\frac{\vec l}{l} \cdot (- \vec{q} + 
\frac{\omega_q}{m_N}\vec{p}\; )\Big] 
  R_{^3\!P_0,\; p_i}(l)~\}~\label{piLH}.
\ea 
Here $p=|\vec p\; |$, and
$x$ denotes the cosine of the angle 
between $\vec p$ and $\vec q$, 
i.e., $\vec p\cdot\vec q=pqx$;  
$\vec l\equiv \vec p+\vec q/2$, 
and the momentum 
$p_c$ is chosen to lie between $p_f$ and $p_m$, 
where $p_m$
is the solution of the equation 
$p_i=\sqrt{p_m^2+p_mqx+q^2/4}$ 
for a given value of $x$. 
The merit of dividing the p-integration range
in the two intervals is that
each $p$-space integral in Eq.(\ref{piLH})
contains only one principal-value part 
coming from either the initial or the final 
NN relative-motion propagator. 
For instance, 
in the second integral of Eq.(\ref{piLH}) the final
state radial wave function takes the following simple form
\ba
 R_{^1\!S_0,\; p_f}(p)=\cos [\delta(^1\!S_0)]m_N 
\frac{K_{\!^1\!S_0} (p,p_f)}{p_f^2-p^2}~, \nonumber
\ea
where $p_f (< p_c)$
is the final-state on-shell momentum.
We evaluate the first term in Eq.(\ref{piLH}) 
directly in $p$-space. 
The second integral in Eq.(\ref{piLH}) 
is calculated 
using a modified Bessel transformation 
outlined in the appendix. 

\vspace{3mm}

Since the above-described method 
for carrying out the r-space calculation of the
1-body amplitude with $\vlk$ is somewhat involved,
it seems safer to check its validity
using some pilot calculation.
If in deriving ${\cal{T}}$ 
we assume $\vec{q}=0$ 
(the ``$\vec{q}=0$ approximation"), 
the evaluation of the transition matrix
element is drastically simplified, and even with
$\vlk$ we need not resort 
to the above lengthy prescription.
We therefore consider it informative
to compare the results of calculations 
with and without the ``$\vec{q}=0$ approximation".
This comparison is given in Appendix B.

\section{Numerical Results }

We now present
the $pp\rightarrow pp\pi^0$ cross sections
calculated with the use of $\vlk$~\cite{vlk01}
as well as typical bare NN interactions.
We consider here three examples
of the bare potential:
Argonne V18~\cite{wir95},
Bonn-B~\cite{Bonnb}, and CD-Bonn~\cite{CDBonn}.
The results for Argonne V18
are taken from Ref.~\cite{SLMK97},
while those for Bonn-B and CD-Bonn
have been newly calculated in the present work. 
It is informative to study the individual behavior
of the impulse- and rescattering-term contributions
before discussing the behavior of 
their combined contributions.

\subsection{Contribution of
the impulse-approximation term}

\begin{figure}
\centerline{\epsfig{file=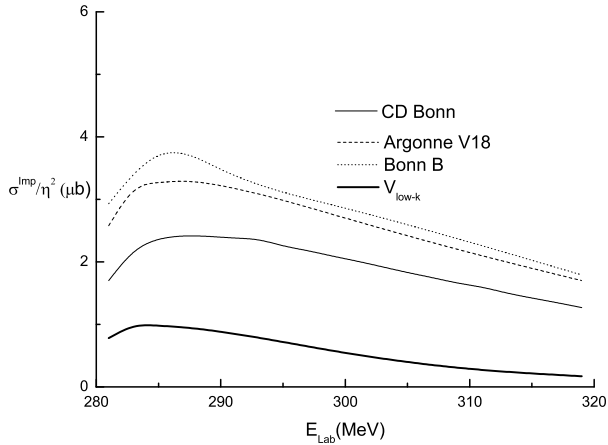,width=8.0cm}} 
\caption{\small
$\sigma^{\rm Imp}$, the 
$pp\rightarrow pp\pi^0$ cross section  
calculated with the impulse term alone. 
The cross section
is given in units of $\eta^2$,
where the ``velocity" $\eta$ 
is defined as the maximum pion momentum 
at a given laboratory energy $E_{Lab}$
of the incident proton divided by the pion mass, 
$m_\pi=135$ MeV. 
}\label{imp}
\end{figure}
We first consider $\sigma^{\rm Imp}$,
the total $pp\rightarrow pp\pi^0$
cross section calculated
with only the impulse-approximation amplitude retained;
{\it viz.,} in evaluating Eq.(\ref{eq:sigmaWb}),
we replace 
$<\!\!p_f[L_fS_f]J_f\,||\,{\cal T}_{l_\pi}(q)
\,||\,p_i [L_iS_i]J_i\!\!>$
with 
$<\!\!p_f[^1\!S_0]\,||\,
{\cal T}^{\rm Imp}_{l_\pi=0}(q)
\,||\,p_i [^3\!P_0]\!\!>$.
Fig.~\ref{imp} shows $\sigma^{\rm Imp}$
as a function of $E_{\rm Lab}$,
the incident proton energy
in the laboratory system.
We note that, 
for the three representative bare NN-potentials,
$\sigma^{\rm Imp}$ varies
up to 40 \%.
These variations are a measure of ambiguity
inherent in a calculation that uses
a bare NN potential.
The fact that the short-distance behavior
of bare NN potentials is not controlled
with sufficient accuracy
underlies this instability.
Fig.~\ref{imp} indicates that 
the use of $\vlk$ leads to a value
of $\sigma^{\rm Imp}$
that is significantly smaller 
(by a factor of 3 or more)
than those for the bare potentials.
A plausible explanation of this difference
is as follows. 
In the one-body transition diagram
[Fig.~\ref{impres}(a)],
the large momentum transfer 
($p\sim \sqrt{m_\pi m_N}$)
between the two nucleons 
needs to be mediated 
by the N-N potential.\footnote{
As pointed out in Ref.\cite{SLMK97},
with the use of a bare NN potential,
one needs to take the upper limit
of $p$-integration very high 
(up to $p \sim 2$ GeV/c) 
before the integral starts to 
show a sign of convergence
for both the impulse and rescattering amplitudes.}
Now, by construction, $\vlk$ only contains 
momentum components below $\Lambda =2\, {\rm fm}^{-1}$, 
whereas the bare potentials carry
very high momentum components (albeit 
in a rather arbitrary manner).
We can expect
that the absence of those high-momentum components
in $\vlk$ suppresses the contribution 
of the one-body transition diagram.

\vspace{3mm}

As mentioned, we are using in the present work
the $\vlk$ derived from the CD Bonn potential.
It is known, however, 
that, so long as one starts from 
a bare potential that belongs to the category of 
modern high-precision phenomenological potentials,
the resulting $\vlk$ is practically model-independent
in the sense that the 
half-on-shell $K$-matrices corresponding
to different bare potentials
are nearly indistinguishable~\cite{vlk01,vlk03}.
This means that $\sigma^{\rm Imp}$
calculated with $\vlk$ corresponding to 
any realistic bare potential
would lie close to the solid line 
in Fig.~\ref{imp}.
Thus the use of $\vlk$ results in
a significant reduction of model dependence
in our calculation.
\begin{figure}
\centerline{\epsfig{file=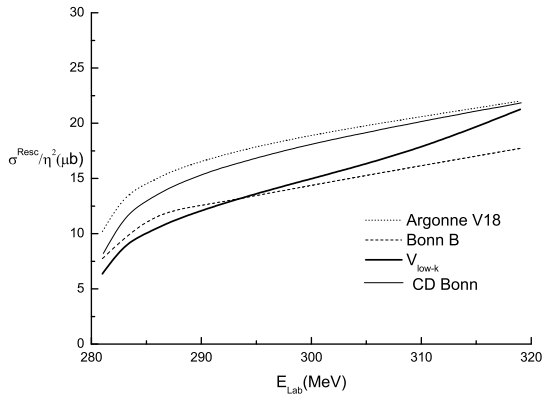,width=8.0cm}} 
\caption{\small
$\sigma^{\rm Resc}$, 
the $pp\rightarrow pp\pi^0$
cross sections calculated with the 
rescattering term alone.
For the explanation of the quantity $\eta$,
see the caption of Fig.1.
}\label{res}
\end{figure}

\vspace{3mm} 
\subsection{Contribution of the rescattering term}

Fig.~\ref{res} gives $\sigma^{\rm Resc}$,
the total $pp\rightarrow pp\pi^0$
cross section calculated
with only the rescatttering term contribution retained;
{\it i.e.,} in evaluating Eq.(\ref{eq:sigmaWb}),
we replace 
$<\!\!p_f[L_fS_f]J_f\,||\,{\cal T}_{l_\pi}(q)
\,||\,p_i [L_iS_i]J_i\!\!>$
with 
$<\!\!p_f[^1\!S_0]\,||\,
{\cal T}^{\rm Resc}_{l_\pi=0}(q)
\,||\,p_i [^3\!P_0]\!\!>$.
The figure indicates that,
for the three different choices 
of the bare NN-potential,
$\sigma^{\rm Resc}$ shows variations of 
about 30 \%, 
while the use of $\vlk$ leads to
$\sigma^{\rm Resc}$ that lies 
more or less within the range of these variations.
In the rescattering diagram [Fig. 1(b)],
a substantial fraction of
the momentum transfer between the two nucleons
can be carried by 
the exchanged pion, and therefore the NN interactions
need not directly support 
a large momentum transfer.
This feature explains why the change
in $\sigma^{\rm Resc}$ is less pronounced
than $\sigma^{\rm Imp}$ as we switch 
from the bare potentials to $\vlk$.
It is worth re-emphasizing here
that, although $\sigma^{\rm Resc}(\vlk)$ 
in Fig.~\ref{res} was obtained with the 
$\vlk$ derived from the CD-Bonn potential,
the result should be considered model-independent
in the sense discussed in the preceding subsection.

\subsection{Combined contributions of 
the impulse-approximation and rescattering terms}

\begin{figure}
\centerline{\epsfig{file=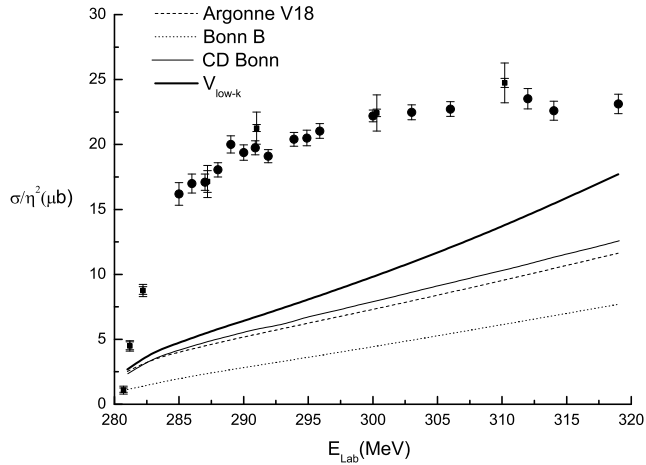,width=9.5cm}} 
\caption{\small
The total cross sections calculated from
the full amplitude
consisting of the impulse and rescattering terms.
The experimental data points are taken 
from Ref.~\cite{meyetal90}
(solid circles) and Ref.~\cite{Uppsala} (solid squares).
For the explanation of the quantity $\eta$,
see the caption of Fig.1.
}\label{tot}
\end{figure}

We now consider 
the total $pp\rightarrow pp\pi^0$ cross section,
$\sigma$, 
calculated with the full transition amplitude
consisting of the one-body and two-body terms;
thus $\sigma$
is obtained from Eq.(\ref{eq:sigmaWb})
with the transition amplitude given by
Eqs.~(\ref{eq:Treldecomp}), (\ref{eq:TImppspace}), 
(\ref{eq:TRescpspace}).
The results are shown in Fig.~\ref{tot}
for the three choices of
the bare potential and for $\vlk$,
along with the experimental values of
$\sigma_{pp\rightarrow pp\pi^0}$. 
Fig.~\ref{tot} indicates that
the use of $\vlk$ leads to a rather visible enhancement 
of $\sigma$ over the results obtained
with the bare potentials.
This enhancement is related 
to the suppression of the impulse-approximation amplitude 
corresponding to $\vlk$. 
As pointed out in the earlier $\chi$PT 
calculations~\cite{pmmmk96, cfmv96},
the impulse and rescattering amplitudes
tend to interfere destructively,
and in the case of a bare NN interaction
the cancellation between the two amplitudes
is quite substantial, leading to a significantly suppressed 
value of $\sigma$ as compared with 
the individual magnitudes
of $\sigma^{\rm Imp}$ 
and $\sigma^{\rm Resc}$.
The smaller impulse-approximation amplitude obtained 
with the use of $\vlk$ somewhat
diminishes the extent
of this destructive interference,
resulting in a larger value of $\sigma$.
The enhancement of the cross
section obtained with $\vlk$ brings 
the calculated values of $\sigma$ closer
to the experimental values.
It is to be noted,
however, that the energy dependence 
of $\sigma$ obtained with $\vlk$ 
differs significantly from  
the experimentally observed behavior. 
We remark once again that the 
$\sigma(\vlk)$ shown in Fig.~\ref{tot} 
should be essentially independent
of the choice of a bare potential from which
$\vlk$ is derived (see subsection 4.1).
%
%
\begin{figure}
\centerline{\epsfig{file=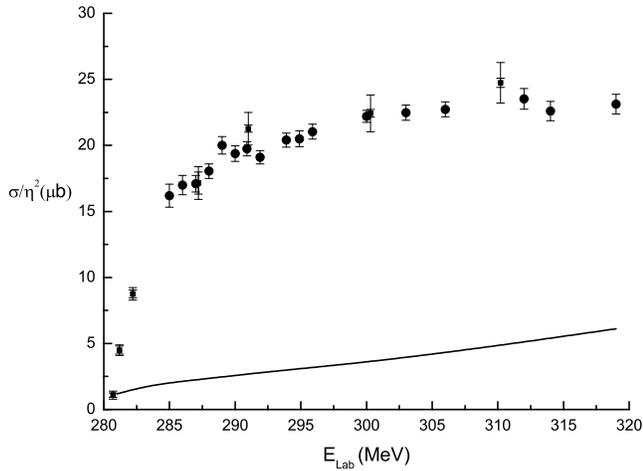,width=9.5cm}} 
\caption{\small
The total cross sections calculated from
the full amplitude
consisting of the impulse and rescattering terms evaluated in the 
{\it fixed kinematics approximation}.
The experimental data points are taken 
from Ref.~\cite{meyetal90}
(solid circles) and Ref.~\cite{Uppsala} (solid squares).
For the explanation of the quantity $\eta$,
see the caption of Fig.1.
}\label{totFKA}
\end{figure}

\vspace{3mm}

The above results correspond to the case where
the four-momentum $k$ of the exchanged pion
is specified according to 
Eqs.(\ref{eq:kchoicea}) and (\ref{eq:kchoiceb}).
As discussed earlier, however, 
there is an argument that favors FKA 
in a certain context~\cite{hmmsv01}.
It therefore seems informative to repeat our calculation
with the use of FKA.
The $\sigma$ corresponding to this case
is shown in Fig.~\ref{totFKA}.
We observe that the cross sections obtained in FKA
are smaller than those in Fig.~\ref{tot}, representing
a larger deviation from the experimental values,
and that the energy dependence of $\sigma$
does not resemble the experimentally observed behavior.

\vspace{3mm}

\begin{figure}
\centerline{\epsfig{file=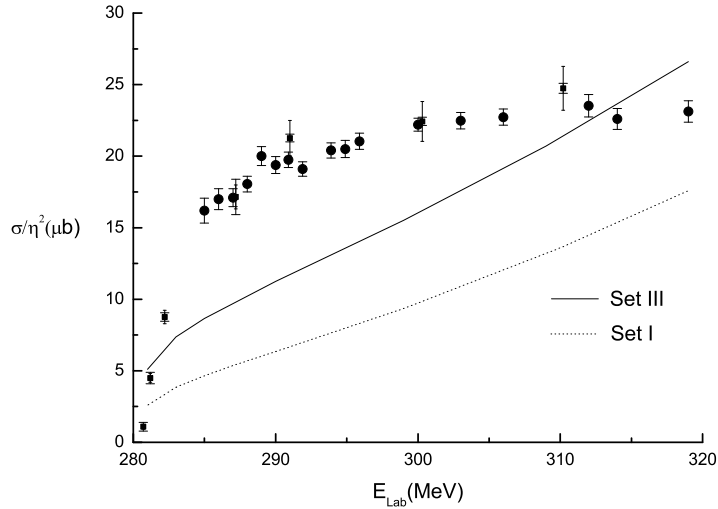,width=9.5cm}} 
\caption{\small
The total cross section $\sigma$
calculated with the full amplitude 
(including both the one-body and two-body amplitudes)
for $\vlk$.
The results obtained with the parameter sets I and III
are compared.
The experimental data points are taken from Ref.~\cite{meyetal90}
(solid circle) and Ref.~\cite{Uppsala} (solid square).
}\label{tot3}
\end{figure}

However, we need to discuss here
the dependence of our results on 
the values of the LEC, $c_1$, $c_2$ and $c_3$.
The above results were
obtained for the ``standard values"
of $c_1$, $c_2$ and $c_3$ 
given in Eq.(\ref{eq:lecoef}).
%
These were originally deduced in Ref.~\cite{bkm95}
and quoted in Ref.~\cite{SLMK97} 
as ``parameter set I".
The allowed ranges of these LECs 
were discussed in Ref.~\cite{SLMK97,bkm97},
where, in addition to the parameter set I,
two more sets were considered
as examples of other possible choices.
For convenience, we tabulate 
these three sets of parameters:
\begin{eqnarray}
\lefteqn{{\rm Parameter}\;\;{\rm set}\;\;{\rm I}}
\nonumber\\
& &\;\;\;\;\;\;\;\;\;c_1=-0.87\;{\rm GeV}^{-1},\;\;
c_2=3.34\;{\rm GeV}^{-1},\;\;
c_3=-5.25\;{\rm GeV}^{-1} 
\label{eq:set1}\\
\lefteqn{{\rm Parameter}\;\;{\rm set}\;\;{\rm II}}
\nonumber\\
& &\;\;\;\;\;\;\;\;\;c_1=-0.87\;{\rm GeV}^{-1},\;\;
c_2=4.5\;{\rm GeV}^{-1},\;\;
c_3=-5.25\;{\rm GeV}^{-1} 
\label{eq:set2}\\
\lefteqn{{\rm Parameter}\;\;{\rm set}\;\;{\rm III}}
\nonumber\\
& &\;\;\;\;\;\;\;\;\;c_1=-0.98\;{\rm GeV}^{-1},\;\;
c_2=3.34\;{\rm GeV}^{-1},\;\;
c_3=-5.25\;{\rm GeV}^{-1} ~. 
\label{eq:set3}
\end{eqnarray} 
To get a measure of the sensitivity 
to the choice of the LECs,
we have repeated our calculation of $\sigma$
for $\vlk$ using the parameter sets II and III.
The results are shown in Fig.~\ref{tot3}
together with those for the set I;
in fact, since the set II gives 
practically the same result as the set I,
we give in the figure only the results 
for the sets I and III.
Fig.~\ref{tot3} indicates
that the set III, which differs
from the set I only by a modest 12\% change
in $c_1=-0.98$ GeV$^{-1}$,
enhances $\sigma$ considerably,
bringing the calculated values of $\sigma$
closer to the experimental values.
However, the energy dependence of 
the theoretical $\sigma$ 
remains dissimilar 
to the experimentally observed behavior.
Figs.~\ref{tot}, \ref{totFKA} and \ref{tot3} 
seem to suggest that,
in order to fully explain the magnitude and
incident-energy dependence of the 
$pp\rightarrow pp\pi^0$ cross sections
near threshold,
one probably needs to include terms 
of chiral orders higher 
than considered here.
We remark in this connection
that the possible importance of 
two-pion exchange diagrams 
in a $\chi$PT calculation 
of the $pp\rightarrow pp\pi^0$ reaction
was pointed out in Refs.~\cite{dkms99, hk02}.

\section{Discussion and Summary 
}

We have carried out a hybrid $\chi$PT calculation
of the cross section $\sigma$
for the $s$-wave pion production reaction,  
$pp\rightarrow pp\pi^0$,
with the use of $\vlk$.
$\vlk$ is a low-energy effective potential
derived from a high-precision
phenomenological potential
(called a ``bare" potential in our context)
by integrating out momentum components 
higher than $\Lambda\sim 2$ fm$^{-1}$.
The results obtained with $\vlk$
are compared with those obtained with the
three representatives bare potentials,
ANL V18, Bonn-B and CD-Bonn.
The principal features of our calculation 
based on $\vlk$ are summarized as follows.\\
(1) A hybrid $\chi$PT calculation based on
a bare potential has the ``momentum mismatch" problem
that the initial and final nuclear
wave functions generated by the bare potential
involve very high momenta,
whereas the transition operators derived
from $\chi$PT can be used only within a limited
momentum range ($p \le $ 300 MeV).
This momentum mismatch problem is significantly 
mitigated by the use of $\vlk$.\\
(2) Reflecting the fact that the short-distance 
behavior of the bare potential is not well controlled,
the $\sigma$'s calculated with the above-mentioned
three bare potentials exhibit $\sim$40 \% variance.
This kind of model dependence
practically disappears with the use of $\vlk$,
since different choices of a bare potential 
lead to practically equivalent $\vlk$'s~\cite{vlk03}. 
This feature allows us to better focus on the question
whether the transition operator for the 
$pp\rightarrow pp\pi^0$ reaction
derived from $\chi$PT up to next-to-leading order
is adequate or not.\\
(3) The calculation with $\vlk$ enhances $\sigma$
over the values obtained with the bare potentials,
and, with certain choices of the relevant LECs,
$\sigma$ can come close to the experimental values 
for some range of the incident energy.
It is however unlikely that the 
magnitude and energy dependence of $\sigma$
can be fully reproduced with the transition operators
considered in this work.
Thus it seems necessary to consider higher-order
transition operators. 

\vspace{3mm}

For formal consistency, 
it is desirable to go beyond the hybrid $\chi$PT approach
by employing N-N potentials derived from $\chi$PT.
This is however a major task relegated 
to the future.
We remark in this connection that,
for reactions that only involve the 
rearrangement of the nucleons,
there has been much progress in constructing 
a framework that is formally consistent 
with effective field theory~\cite{Epelbaum,BvK02}. 

\vspace{3mm} 

A related issue is that we concentrated here on the 
consequences of changing the 
nuclear wave functions from those generated by the bare 
NN potentials to those generated by $\vlk$, 
{\it without} taking account of the 
possible renormalization of the transition operators 
due to the truncation of model space.\footnote{For 
discussion of
some formal aspects of the use  of 
$\vlk$ in hybrid $\chi$PT,
see Ref.~\cite{nakamura04}.}
As is well known, a reduction of nuclear model space 
in general entails a corresponding modification of 
operators for the nuclear observables. 
Again, to fully address this issue, we need 
to go beyond the hybrid 
$\chi$PT used in this work. 
In the present study of exploratory nature,
we have concentrated on a hybrid $\chi$PT evaluation 
of the transition operators that arise from the tree diagrams. 
At this chiral order there are no loop corrections
to the LECs ($c_i$, $i$ = 1, 2, 3) 
and the other coefficients appearing  in 
${\cal L}_{ch}$ like, e.g. $g_A$.
We may therefore expect that, 
although the scale of $\chi$PT ($\Lambda_\chi \sim 1$ GeV)
is larger than the momentum cutoff scale 
($\Lambda \sim 2$ fm$^{-1}$) 
used in deriving $\vlk$,
this difference does not lead to a drastic 
renormalization of the transition operators.  
To turn around the argument,
the issue of the renormalization of the transition operators
is coupled with the treatment of higher chiral-order terms,
and these two aspects need to be addressed 
simultaneously. 
This important question, however, is beyond the scope of
our present study, which is limited to the tree diagrams.

\vspace{3mm} 

As an immediate follow-up of the work described here,
we are studying~\cite{ykms05}
the expected important contributions 
from the two-pion exchange diagrams~\cite{dkms99, hk02}
in a hybrid $\chi$PT calculation with  $\vlk$.

\section{Acknowledgment}
We are grateful to T.T.S. Kuo 
for generously allowing us to use 
the $\vlk$ code. We are also grateful to R. Machleidt
for providing the USC group 
with a code to test the Bonn-B potential, 
and also for sending the CD-Bonn code used 
in this work to the Osaka University group. 
We are appreciative of the insightful comments we 
received from S. Ando and C. Hanhart.  
This work is supported in part by the US National Science
Foundation, Grant Nos. PHY-0140214 and PHY-0457014. 
TS gratefully acknowledges financial support from the 
Japan Society for the Promotion of Science, 
Grant-in-Aid for Scientific Research (C) 15540275. 

\section*{Appendix A}

With the use of $\vlk$,
the numerical evaluation 
of the second integral of 
the impulse amplitude [Eq.(\ref{piLH})] 
requires the ``modified" Bessel transformation, 
\begin{eqnarray}
R_{(LS)J,p_0}(r) = \sqrt{\frac{2}{\pi}}\,i^L
\!\int_0^\Lambda \! p^2dp\,j_L(pr)R_{(LS)J,p_0}(p)\, . 
\label{eq:Bessel2}
\end{eqnarray} 
However, since $\int_0^\Lambda 
j_L(pr)j_L(pr^\prime) p^2dp
\neq\frac{\pi}{2r^2}\delta (r-r^\prime)$,  
the inverse transformation for Eq.(\ref{eq:Bessel2})
gets complicated, which presents us from
arriving at an r-space expression 
similar to Eq.(\ref{eq:TImprspace}). 
We therefore replace  
$ R_{(LS)J,p_0}(p)$ in Eq.(\ref{eq:Bessel2})
with the K-matrix expression Eq.(\ref{eq:Kmatrix})
and then we use in Eq.(\ref{piLH})
the Bessel transformation
\begin{eqnarray}
R_{(LS)J,p_0}(k) = \sqrt{\frac{2}{\pi}}\,i^{-L}
\!\int_0^\infty \! r^2dr\,j_L(kr)R_{(LS)J,p_0}(r)\, ,  
\end{eqnarray} 
where  $k \le \Lambda $ is understood. 
In our numerical evaluation of the impulse 
amplitude with $\vlk$, 
only the initial wave function in 
the second integral of Eq.(\ref{piLH}) is evaluated 
using the above prescription.

\section*{Appendix B}

\begin{figure}
\centerline{\epsfig{file=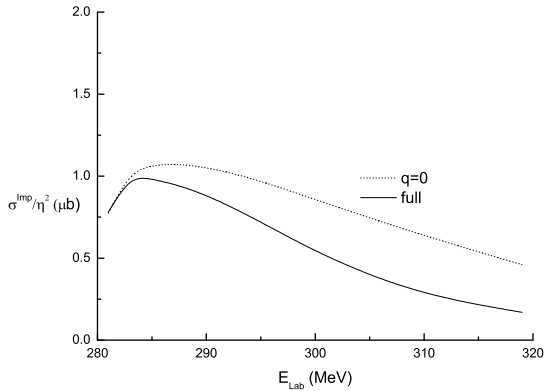,width=8.0cm}} 
\caption{\small
$\sigma^{\rm Imp}$ calculated 
for $\vlk$ without or with 
the $\vec{q}=0$ approximation.
The solid (dotted) line corresponds to the  
calculation without (with)
$\vec{q}=0$ approximation. 
For the explanation of the quantity $\eta$,
see the caption of Fig.1.
}\label{impq0}
\end{figure}

To check the validity of the numerical
techniques described in the text,
it is informative to compare 
the $\sigma^{\rm Imp}$ calculated for $\vlk$
using the prescription explained
in subsection 3.2 with
the $\sigma^{\rm Imp}$ obtained
in the ``$\vec{q}=0$ approximation",
wherein the transition operators are derived
under the simplifying assumption
that the out-going pion has no momentum,
$\vec{q}=0$.
[In evaluating the phase space, we do
treat $\vec{q}$ as a variable.]
In the $\vec{q}=0$ approximation, 
the impulse term can be evaluated
in p-space in a straightforward manner
without any complicated handling 
of the principal value integrals.
The results for the two cases,
with and without the $\vec{q}=0$ approximation, 
are shown in Fig.~\ref{impq0},
and we compare this figure with Fig. 2 in 
Ref.~\cite{SLMK97}, 
which presents $\sigma^{\rm Imp}$ calculated 
for the bare potentials with and without 
the $\vec{q}=0$ approximation.
According to Fig. 2 in Ref.~\cite{SLMK97},
$\sigma^{\rm Imp}$ obtained in the $\vec{q}=0$ approximation
tends to become somewhat larger 
than  $\sigma^{\rm Imp}$
obtained without using the $\vec{q}=0$ approximation. 
The fact that Fig.~\ref{impq0} exhibits a similar tendency
may be taken as an indication 
that the somewhat lengthy procedure
we use in handling the principal-value integrals
for the $\vlk$ case is reliable.

\end{document}